\documentclass[aip,rsi,amsmath,amssymb,longbibliography,superscriptaddress,reprint,nobibnotes]{revtex4-2}

\usepackage{graphicx}
\usepackage{amsmath}
\usepackage{upgreek}
\usepackage{color}
\usepackage{xcolor}
\definecolor{navy}{rgb}{0,0,0.64}
\usepackage{setspace}
\usepackage{hyperref}
\hypersetup{colorlinks,allcolors=black}
\usepackage{nccmath}
\usepackage{amssymb}
\usepackage{bm}
\usepackage{lineno} 
\usepackage{titlesec}
\usepackage[utf8]{inputenc}
\usepackage[T1]{fontenc}
\usepackage{booktabs}
\usepackage{array}
\usepackage{multirow}

\usepackage{tikz}
\DeclareRobustCommand\circled[1]{\tikz[baseline=(char.base)]{\node[shape=circle,draw,inner sep=0.4pt] (char) {#1};}}

\setcitestyle{super}

\newcommand{\papertitle}{A table-top few-femtosecond broadband extreme-ultraviolet absorption spectrometer with cryogenic cooling}

\newcommand{\Cal}{Department of Chemistry, University of California at Berkeley, Berkeley, CA 94720, USA}
\newcommand{\LBNL}{Materials Sciences Division, Lawrence Berkeley National Laboratory, Berkeley, California 94720, USA}
\newcommand{\CalPhys}{Department of Physics, University of California at Berkeley, Berkeley, CA 94720, USA}

\newcommand{\Stanford}{Departments of Physics and of Applied Physics, Stanford University, Stanford, CA 94305, USA}

\makeatletter
\def\@email#1#2{%
 \endgroup
 \patchcmd{\titleblock@produce}
  {\frontmatter@RRAPformat}
  {\frontmatter@RRAPformat{\produce@RRAP{*#1\href{mailto:#2}{#2}}}\frontmatter@RRAPformat}
  {}{}
}%
\makeatother

\begin{document}

\title{\papertitle}

% \begin{center}
% \textbf{\large \papertitle}
% \vspace{0.3cm}

% Authors$^{1,\,2,\,{\color{navy}\ast}}$

% \vspace{0.5cm}
% (Dated: \today)
%\end{center}

% \vspace{0.23cm}
% \begin{small}
% \begin{singlespace}
% {\it
% \noindent$^1$\Cal
% \vspace{0.23cm}
% }
% \end{singlespace}
% \end{small}

\author{Sheng-Chih Lin\textcolor{navy}{$^\dagger$}}
\affiliation{\Cal}
\affiliation{\LBNL}
\author{Alfred Zong\textcolor{navy}{$^\dagger$}}
\affiliation{\Cal}
\affiliation{\LBNL}
\affiliation{\Stanford}
\author{Emma Berger}
\affiliation{\Cal}
\affiliation{\CalPhys}
%\author{Dmitry Lebedev} %sample grower
%\affiliation{\NWU}
%\author{Thomas Wei Song} %sample grower
%\affiliation{\NWU}
\author{Bailey R. Nebgen}
\affiliation{\Cal}
\affiliation{\LBNL}
\author{Marcus Hui}
\affiliation{\Cal}
\author{Shuaiwei Pan}
\affiliation{\Cal}
\affiliation{\LBNL}
\affiliation{\CalPhys}
\author{Jackson McClellan}
\affiliation{\Cal}
\affiliation{\LBNL}
%\author{Mark C. Hersam} %sample grower
%\affiliation{\NWU}
\author{Michael W. Zuerch}
\affiliation{\Cal}
\affiliation{\LBNL}
\email[]{Correspondence to M.W.Z.: mwz@berkeley.edu~~          $^\dagger$These authors contributed equally: S.-C.L. and A.Z.}

\date{\today}

\begin{abstract}
We present a table-top cryogenic ultrafast broadband XUV absorption spectroscopy (c-UBXAS) beamline designed for temperature-dependent and time-resolved investigations of quantum materials. The instrument combines a broadband high-harmonic generation source spanning 22–73~eV with automated image registration and cryogenic sample control down to 20~K, enabling element-specific measurements under both equilibrium and nonequilibrium conditions. The beamline provides sub-50-meV energy resolution and a sub-10-fs instrument response function, while maintaining long-term stability suitable for extended, ultrasensitive measurements. Benchmark experiments on NiI$_2$, a van der Waals multiferroic, reveal temperature-dependent spectral evolution across its magnetic phase transitions, demonstrating the ability to identify element-specific contributions to the ground state and its transient response in the femtosecond regime. This instrument introduces the opportunity to unravel the very first response of a solid-state material during light–matter interaction, paving the way for understanding the complex phase space of nonequilibrium dynamics and emergent states in strongly correlated systems where the few-femtosecond electronic response has eluded most other types of time-resolved techniques.
\end{abstract}

\maketitle

%Strongly correlated materials exhibit a rich variety of emergent phenomena arising from the interplay of charge, spin, orbital, and lattice degrees of freedom. Understanding the microscopic mechanisms 

\begin{figure*}[htb!]
	\includegraphics[width=\textwidth]{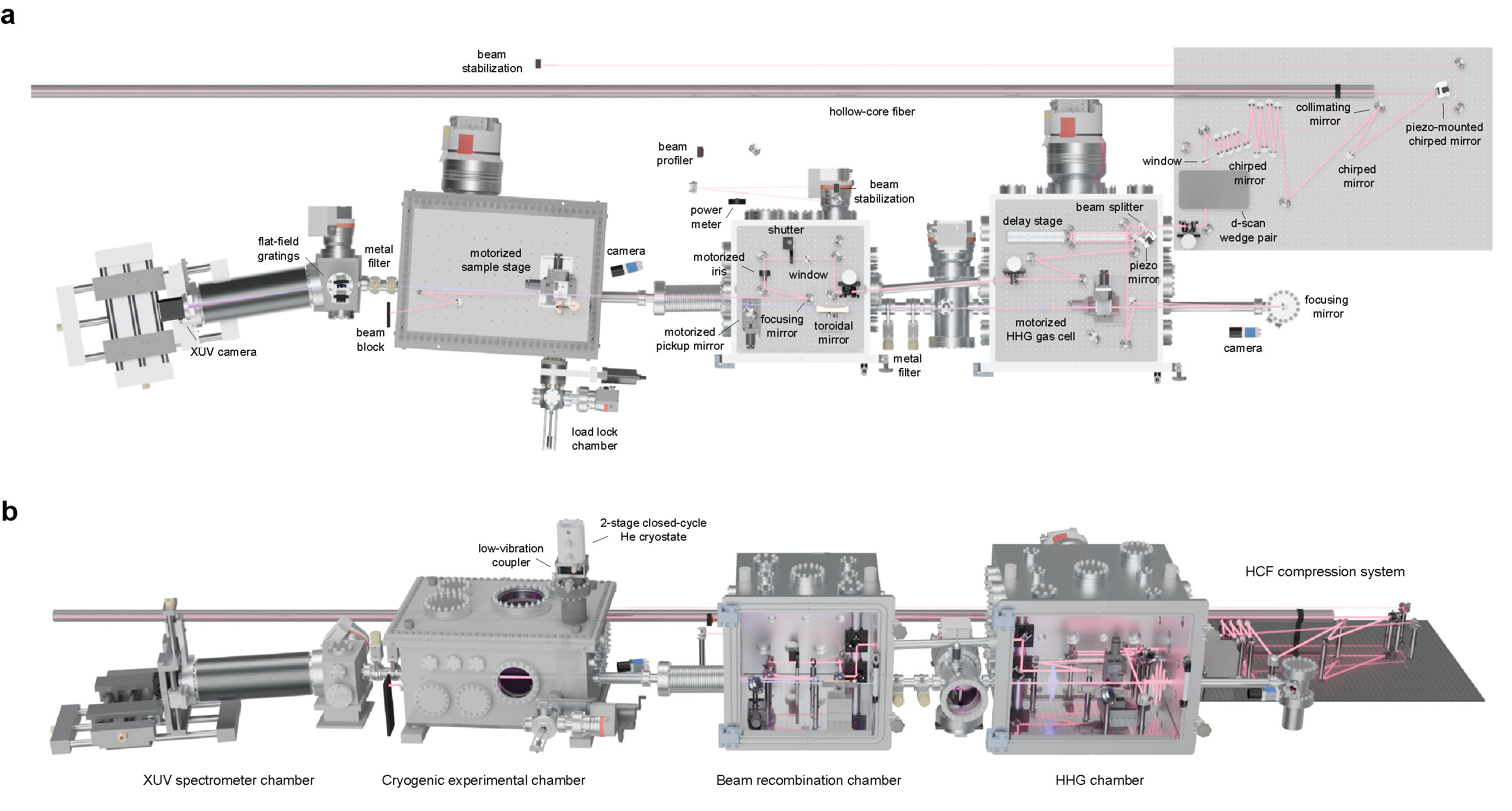}
	\caption{\textbf{Overview of the cryogenic ultrafast broadband XUV absorption (c-UBXAS) beamline.} \textbf{a},~Top-view rendering of the beamline showing the optical layout and major components, including the hollow-core-fiber (HCF) compression system, high-harmonic generation (HHG) chamber, beam recombination chamber, cryogenic experimental chamber, and XUV spectrometer. \textbf{b},~Side-view rendering of the complete beamline assembly highlighting the relative arrangement of the major vacuum chambers and optical subsystems.}
\label{fig:1}
\end{figure*}

%\newpage
\section{Introduction}
Quantum materials host a rich variety of emergent phenomena, such as superconductivity \cite{Orenstein2000, Puphal2026}, (multi)ferroicity \cite{Catalan2012,Fiebig2016}, and charge-density-wave order \cite{Rossnagel2011}, providing fertile ground for both fundamental scientific discovery and next-generation technologies \cite{Tokura2017, Bloch2022, Basov2017}. At the microscopic level, these collective behaviors originate from electronic states shaped by both their constituent atomic orbitals and the geometric configuration of the constituent atoms. %At the microscopic level, these collective behaviors originate from the interplay between electronic states derived from their constituent atomic orbitals and coordination environment. 
By monitoring transitions from localized core states into unoccupied valence states, X-ray absorption spectroscopy (XAS) provides element- and orbital-specific information on the local electronic structure and bonding environment, making it a powerful tool for investigating the origins of these emergent material properties \cite{DeGroot2008}. Moreover, the recent extension of XAS into the sub- to few-femtosecond (fs) regime has enabled the investigation of transient carrier evolutions with distinct hallmarks for electrons and holes \cite{Zurch2017, Molesky2021, Cushing2020}, photoinduced structural change including coherent atomic motions \cite{Geneaux2021b, DeVos2025, Lauren2025}, as well as elementary many-body dynamics such as screening, charge transfer, and electron-electron scattering, which collectively govern the formation and dynamics of emergent quantum phases in solids \cite{Petek1997,Bovensiepen2012,Xu2025,Duris2020,Guo2024,Summers2023}. By combining elemental and orbital specificity with this ultrafast temporal resolution, time-resolved XAS therefore provides a unique window into coupled electronic and structural degrees of freedom by resolving contributions from individual constituent elements that are often difficult to separate in conventional optical spectroscopies.

Despite their transformative capabilities, XAS measurements remain dependent on large-scale user facilities, which motivates the development of laboratory-scale setups that offer greater accessibility and experimental flexibility. To this end, high-harmonic generation (HHG) has emerged as a powerful tabletop alternative \cite{Krausz2009,Lloyd-Hughes2021,Zong2023}. The resulting broadband extreme-ultraviolet (XUV) pulses possess both ultrashort durations and broad spectral coverage, enabling simultaneous access to multiple elemental absorption edges within a single measurement. Over the past two decades, HHG-based XUV spectroscopy has been successfully applied to investigate phenomena ranging from strong-field and photochemical dynamics in gaseous \cite{Loh2008, Kobayashi2018, Kobayashi2019, Drescher2025} and liquid systems \cite{smith2020}, to excitonic \cite{Chang2021, Quintero-Bermudez2024, Geneaux2020, Gannan2025}, magnetic \cite{La-O-Vorakiat2009, Siegrist2019, Ash2023, Johnsen2023}, and carrier dynamics \cite{Schultze2014, Cushing2020, Molesky2021, Zurch2017, Neb2026} in condensed-matter systems. While these studies have established ultrafast XUV spectroscopy as a powerful probe of nonequilibrium phenomena, its application to phase-dependent behavior in quantum materials remains relatively underexplored \cite{Jager2017, Heinrich2023}. This is in part because many collective phases of interest manifest through subtle temperature-dependent modifications of electronic structure, making precise temperature control essential for establishing the relationship between equilibrium electronic structure, emergent order, and nonequilibrium dynamics. However, ultrafast XUV spectroscopy platforms capable of temperature-dependent studies are still rare.

Here, to address the need for temperature control, we have developed a cryogenic ultrafast broadband XUV absorption spectroscopy (c-UBXAS) beamline commissioned at UC Berkeley. The platform enables temperature-dependent static and ultrafast XUV measurements of solid-state materials from 20 to 350~K while maintaining broadband spectral coverage and femtosecond temporal resolution. We present the design and characterization of the beamline, detailing the functionality and implementation of each major subsystem, including the HHG source, beam recombination chamber, cryogenic measurement chamber, and XUV spectrometer. Automated temperature-dependent data acquisition further enables efficient characterization of electronic structure across phase transitions. As a representative demonstration, we perform static and ultrafast XUV measurements on a van der Waals multiferroic NiI$_2$, simultaneously probing the I $N_{4,5}$ and Ni $M_{2,3}$ absorption edges. The measurements illustrate the ability of the platform to connect temperature-dependent electronic structure across phase transitions with element-specific photoinduced dynamics.

\section{Technical description}

\subsection{Overview of instrument}
A schematic diagram of the cryogenic broadband XUV beamline is shown in Fig.~\ref{fig:1}. The instrument is built around a commercial Ti:sapphire amplifier system (Coherent Legend Elite Duo HE) delivering 35~fs near-infrared (NIR) pulses centered at 790~nm with pulse energies up to 13~mJ at a repetition rate of 1~kHz. Approximately 5~mJ of the amplifier output is utilized for beamline operation. The entire beamline is mounted on a vibration-isolated optical table equipped with automatic re-leveling pneumatic vibration isolators (Newport S-2000A Stabilizer) to minimize mechanical vibrations and maintain long-term alignment stability. The beamline consists of five major subsystems: a hollow-core-fiber (HCF) compression system, a HHG chamber, a beam recombination chamber, a cryogenic experimental chamber, and an XUV spectrometer chamber (Fig.~\ref{fig:1}b). Vacuum throughout the beamline is maintained by a two-stage pumping architecture comprising dry roughing pumps (Anest Iwata) backing turbomolecular pumps (Pfeiffer Vacuum), providing operating pressures ranging from high vacuum in the HHG source region to ultrahigh vacuum in the experimental and spectrometer chambers. Each subsystem is described below in the order encountered by the laser pulse as it propagates through the beamline.

\subsection{HCF compression system}
\begin{figure}[tbh!]
	\includegraphics[scale=0.6]{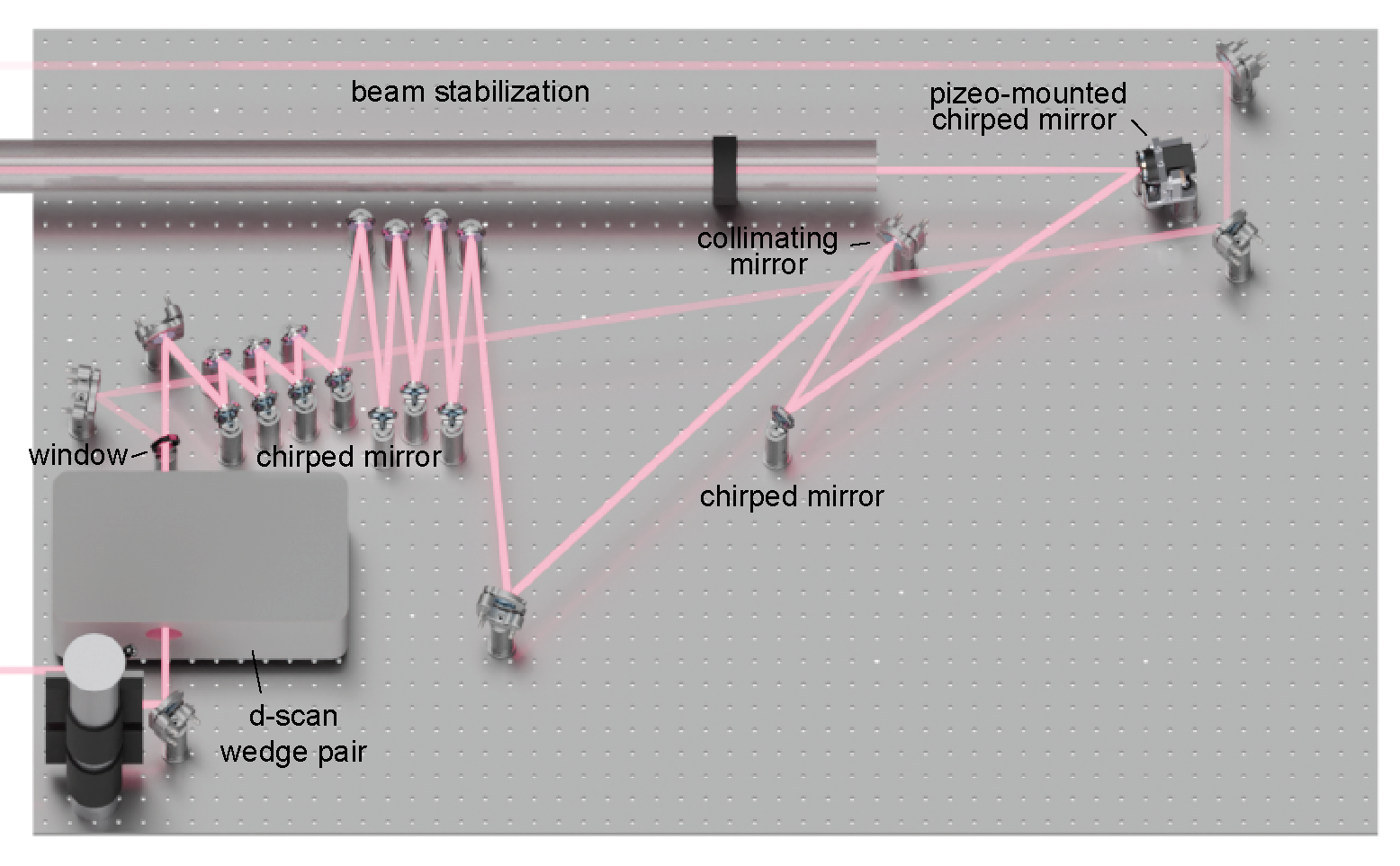}
	\caption{\textbf{Expanded view of the hollow-core-fiber (HCF) compression system.} The spectrally broadened output is compressed using chirped mirrors and a fused-silica wedge pair, while the pulse duration is characterized by the d-scan diagnostic. Active beam-pointing stabilization is implemented before delivery of the compressed beam to the high-harmonic generation chamber. Unlabeled reflective optics correspond to protected silver mirrors (PF10-03-P01, Thorlabs Inc.) used for beam steering.}
\label{fig:2}
\end{figure}

The HCF compression system is used to spectrally broaden the output of the Ti:sapphire amplifier and generate few-cycle pulses for high-harmonic generation and transient experiments. Figure~\ref{fig:2} illustrates the optical layout of the compression region. To begin, 5~mJ NIR pulses centered at 790~nm are focused into a gas-filled stretched hollow-core fiber (BGB Analytik, 700~$\upmu$m inner diameter and 6.4~m length) for spectral broadening via self-phase modulation. The fiber is statically filled with ultrahigh purity He at a pressure of 20~psi. The spectrally broadened output pulses exhibit a pulse energy of 3~mJ with a root mean squared (rms) pulse-to-pulse fluctuation of $0.21\%$. The output beam is subsequently collimated using a spherical silver mirror ($f=2.5$~m) and temporally compressed by eight pairs of chirped mirrors (PC1332, Ultrafast Innovations) and a fused-silica wedge pair (d-scan, Sphere Ultrafast Photonics), yielding a spectrum spanning approximately 540--1050~nm at $-20$~dB and a typical pulse duration of 3.4~fs (FWHM) \cite{Zong2024}. An active beam stabilization (MRC Systems GmbH) is implemented to maintain the beam pointing at the high-harmonic generation gas cell. The feedback signal is derived from the very weak residual reflection of a 300-$\upmu$m-thick broadband-antireflection-coated fused silica window (Lattice Electro Optics, Inc.) right before the d-scan box. The reflected beam is then relayed onto a quadrant detector located at an equivalent propagation distance from the sampling window to the HHG gas cell (Fig.~\ref{fig:1}a), while beam-pointing corrections are applied through a piezo-actuated steering mirror positioned directly after the fiber exit.

\subsection{HHG chamber}
\begin{figure}[tbh!]
	\includegraphics[width=\columnwidth]{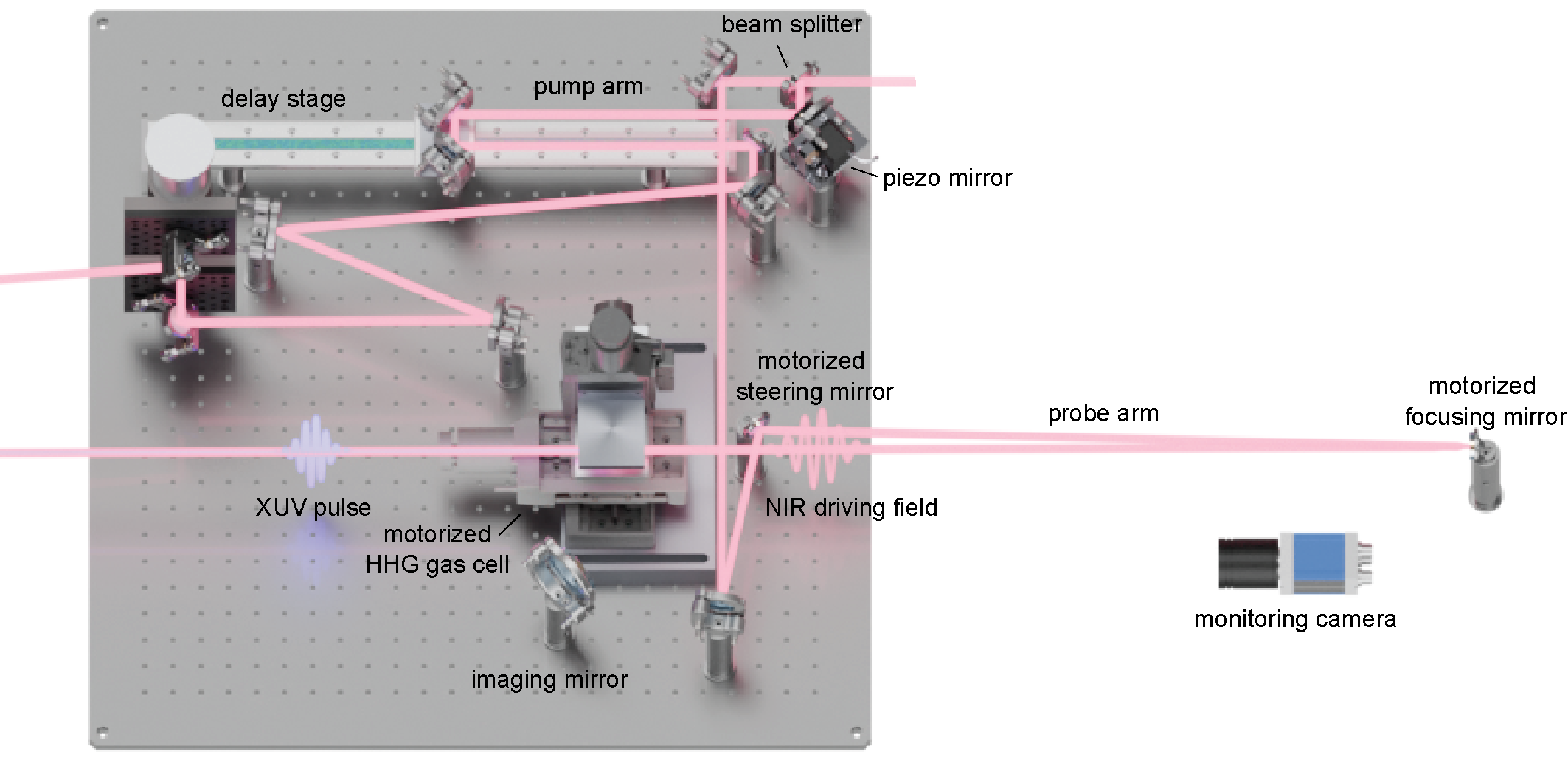}
	\caption{\textbf{Optical layout of the high-harmonic generation (HHG) chamber.} The incoming NIR beam is divided into pump and HHG-driving arms using a custom beam splitter. The pump arm contains a motorized delay stage, while the driving arm is focused into a motorized gas cell for HHG generation. Active beam stabilization and source imaging are implemented for automated beam alignment. The generated XUV pulse and residual NIR driving field are subsequently delivered to the downstream beamline. Unlabeled reflective optics correspond to protected silver mirrors used for beam steering.}
\label{fig:3}
\end{figure}

Following compression, the beam enters the HHG chamber through a 500-$\upmu$m-thick antireflection-coated fused-silica window and is subsequently divided into driving and pump arms with a 90:10 intensity ratio using a custom antireflection-coated beamsplitter (Layertec GmbH). As shown in Fig.~\ref{fig:3}, the pump beam is routed through a vacuum-compatible piezo-actuated linear stage (CLL42, SmarAct Inc.), which controls the relative arrival time between the pump and probe pulses with a 3.3-ns delay range. Locating the delay line inside the vacuum chamber minimizes optical path fluctuations in air, thereby improving long-term time-zero stability during pump-probe measurements. The pump beam is then transported toward the beam recombination chamber.

In the probe arm, both the focusing mirror ($f=0.75$~m) and the steering mirror immediately upstream of it are mounted on piezo-driven tip/tilt stages (STT-25.4, SmarAct; Fig.~\ref{fig:3}). These motorized mounts enable remote optimization of the beam alignment without breaking vacuum. The beam is then focused into a custom-machined Macor ceramic gas cell with a 500-$\upmu$m-diameter circular aperture (Werkzeugbau Simon GmbH) for high harmonic generation. The gas cell is mounted on a motorized three-axis linear stages (Physik Instrumente), allowing remote optimization of phase-matching conditions for the HHG process. The generation medium is supplied through 1/8~inches stainless steel tubing and regulated by a pressure controller (PC-100PSIA-D, Alicat Scientific, Inc.).  A mirror mounted adjacent to the gas cell directs the interaction region to an external alignment camera (Imaging Source), providing real-time visual monitoring of the gas cell position during alignment. To minimize reabsorption of the generated XUV pulses, a differential pumping scheme is implemented. As shown in Fig.~\ref{fig:1}a, two turbomolecular pumps (HiPace~1200 and HiPace~700, Pfeiffer Vacuum) are arranged in series to efficiently remove gas load and to maintain a pressure gradient. The HHG chamber exhibits a base pressure of approximately 7.9~$\times$~10$^{-5}$~Torr during operation. The generated XUV pulses then propagate collinearly with the residual driving field into the downstream beam recombination chamber. 

\subsection{Beam recombination chamber}

\begin{figure}[tbh!]
	\includegraphics[scale=1]{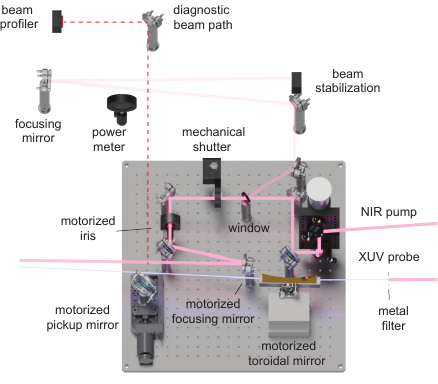}
	\caption{\textbf{Optical layout of the beam recombination chamber.} The NIR pump beam and XUV probe are spatially recombined and directed toward the cryogenic experimental chamber. A motorized toroidal mirror relays the XUV beam, while a motorized focusing mirror controls the pump beam pointing at the sample position. Metal filters suppress the residual NIR driving field transmitted from the HHG chamber. Auxiliary diagnostics, including a beam profiler and a power meter, enable continuous monitoring of the pump beam. A motorized pickup mirror can be inserted into the beam path to redirect the NIR beams on the pump and probe arms to the external diagnostic beam path indicated by the dashed red line for alignment and characterization. A motorized iris and shutter provide remote control of the pump beam delivery. Unlabeled reflective optics correspond to protected silver mirrors used for beam steering.}
\label{fig:4}
\end{figure}

Upon entering the beam recombination chamber, the residual NIR driving field is removed using a metal filter. To accommodate different spectral ranges and experimental requirements, two filter assemblies are mounted on independent manual gate valves (VAT), allowing rapid exchange between different filter materials while maintaining vacuum conditions (Fig.~\ref{fig:1}a). Unless otherwise specified, the measurements presented here were performed using a 200-nm-thick Al filter (Lebow, Co.). The chamber operates at a base pressure on the order of $2.0\times10^{-8}$~Torr. The transmitted XUV pulses are subsequently refocused into the downstream cryogenic experimental chamber by a Ni-coated toroidal mirror (ARW Optical Corporation) with specified tangential and sagittal radii of curvature of 28,656.26~mm and 50.25~mm, respectively. The mirror is operated at $2.4^\circ$ grazing incidence in a $2f$:$2f$ one-to-one imaging configuration, and  is mounted on a motorized four-axis tilt aligner (8071-M, Newport), providing fine optimization of the XUV focus at the sample position. In parallel, the pump beam passes through a mechanical shutter (SH1, Thorlabs) for rapid acquisition of XUV spectra with and without the pump beam. A motorized iris (SID5714, SmarAct) is used for precise control of the pump fluence (Fig.~\ref{fig:4}). The pump beam is subsequently steered and focused onto the sample position by a concave mirror on a two-axis picomotor-actuated mirror mount (8301-V, Newport), allowing fine adjustment of the pump-probe spatial overlap at the sample position. An MRC beam stabilization system is implemented in the pump arm, where the feedback signal is derived from the reflection of a fused-silica window right before the shutter, and the corresponding piezo-actuated mirror is placed in the HHG chamber right after the pump-probe beamsplitter. This beam stabilization maintains long-term pump–probe spatial overlap at the sample position over the entire 3.3~ns delay stage range.

To facilitate the beam characterization and alignment, a motorized translation stage (Physik Instrumente) is positioned downstream of both the toroidal mirror and the pump concave mirror, allowing the pump and probe beams to be redirected to an external diagnostic region; as explained below, the residual NIR beam through the HHG gas cell acts as a proxy for the XUV probe without metal filter insertion. A pyroelectric energy sensor (EnergyMax-RS J-10MB-LE, Coherent) and a beam profiler (S-WCD-UHR, DataRay) positioned at a plane equivalent to the sample position are used to measure the pump pulse energy and beam profile, respectively. This arrangement enables accurate determination of the incident pump fluence. Furthermore, with the metal filter retracted, the NIR driving field can be directed to the diagnostic region. Owing to the shared propagation path and focusing geometry of the NIR driving field and XUV probe, the NIR beam serves as a reliable surrogate for the XUV beam during alignment, facilitating optimization of the pump–probe spatial overlap.

\subsection{Cryogenic experimental chamber}
\begin{figure}[tbh!]
	\includegraphics[width=\columnwidth]{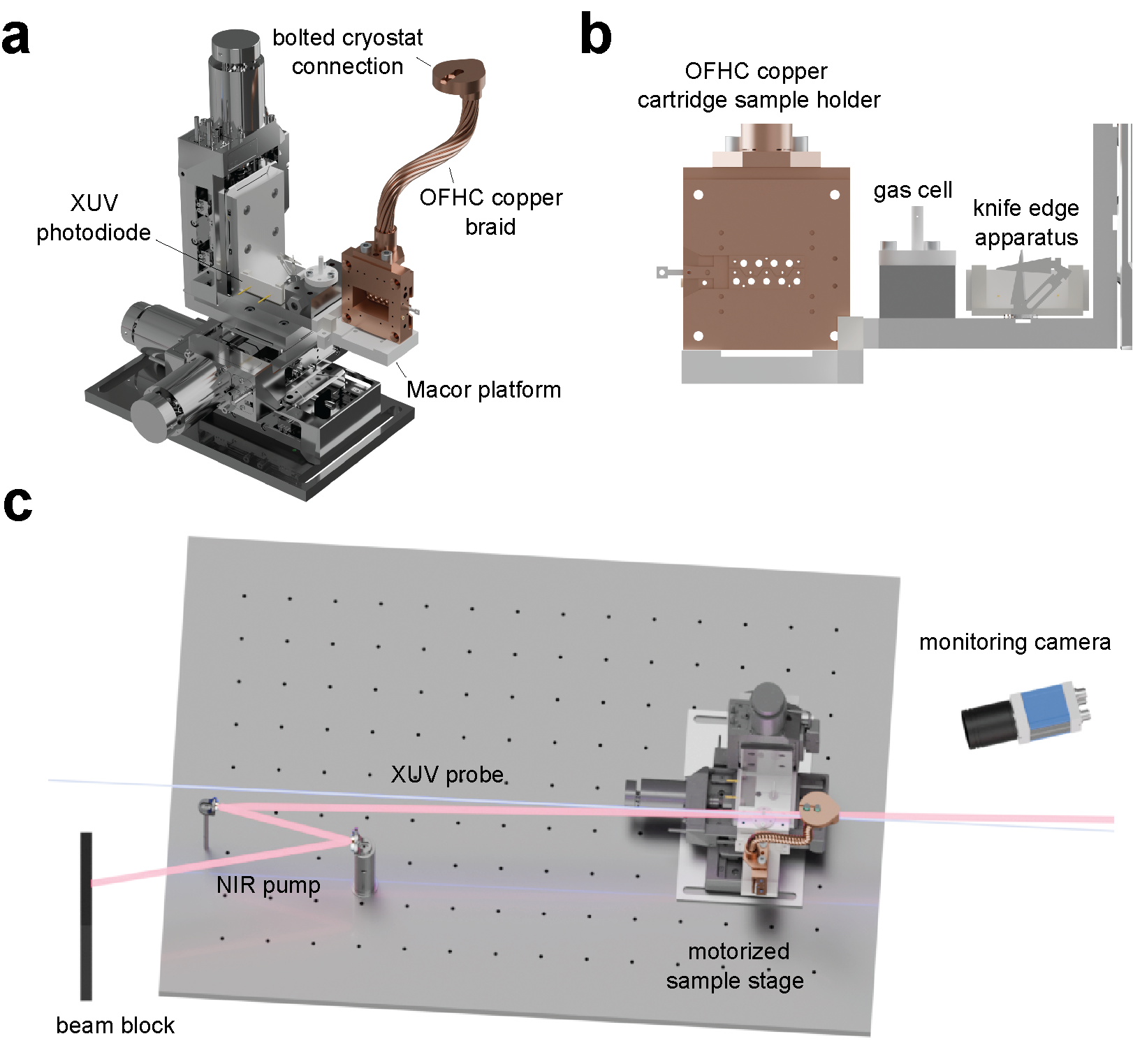}
	\caption{\textbf{Configuration of the cryogenic experimental chamber and sample platform.} \textbf{a},~Three-dimensional rendering of the cryogenic sample platform mounted on the motorized sample stage. The sample holder is fabricated from oxygen-free high-conductivity (OFHC) copper and thermally anchored to the cryostat cold finger through a flexible OFHC copper braid. \textbf{b},~Side view of the sample platform showing the OFHC copper cartridge sample holder together with the auxiliary gas cell and knife-edge apparatus used for energy calibration, temporal characterization, and beam-size measurements. \textbf{c},~Optical layout of the cryogenic experimental chamber. The NIR pump and XUV probe are focused onto the sample mounted on a motorized stage, while the transmitted pump beam is stopped by a beam block.}
\label{fig:5}
\end{figure}

The cryogenic experimental chamber serves as the interaction region for both static and time-resolved XUV spectroscopy measurements. The optical layout of the chamber is shown in Fig.~\ref{fig:5}c. For transient experiments, the XUV probe and NIR pump beams intersect at the sample position with a crossing angle of $\sim4^{\circ}$, and the transmitted pump beam is subsequently steered out of the chamber and directed to a beam dump. The XUV probe is focused to an approximately $100\times100~\upmu\mathrm{m}^2$ spot at the sample position, while the NIR pump beam is expanded to $\sim650~\upmu\mathrm{m}$ to ensure uniform excitation across the probed region. A custom-designed experimental platform mounted on a motorized 3-axis translation stage (Physik Instrumente) accommodates three interchangeable components: a sample holder, a gas cell, and a knife-edge assembly (see  Fig.~\ref{fig:5}a,b), supporting spectroscopy measurements as well as beamline calibration and characterization. 

Solid-state samples are mounted on an oxygen-free high-thermal-conductivity (OFHC) copper holder, which houses a detachable sample cartridge with nine sample slots. This design not only streamlines measurements by providing access to multiple samples within a single vacuum cycle, but also enables sample exchange through a load-lock system without breaking the vacuum of the experimental chamber (Fig.~\ref{fig:1}a). The OFHC copper holder is mounted on the experimental platform through a thermally insulating Macor extension and is coupled to a vibration-isolated closed-cycle liquid $^4$He cryostat (SHI-4XG-UHV-2, ColdEdge Inc.) through a custom 1/2-inch OFHC copper braid assembly with end fittings. Both the sample holder and the copper braid assembly were fabricated by Werkzeugbau Simon GmbH. Sample temperature is regulated using a PID controller (Model~335, Lake Shore Cryotronics), enabling temperature-dependent measurements. The cryostat provides temperature control from approximately 20~K to 350~K, enabling studies across a wide range of thermodynamic phases. While the cryo-cooler is capable of reaching temperatures near 4~K, the minimum sample temperature is limited to approximately 20~K in the current implementation because the radiation shield does not extend to the sample holder. To prevent condensation during cryogenic operation, the experimental chamber is maintained at a base pressure of approximately $1.1\times10^{-9}$~Torr. 

An identical gas cell to that used in the HHG chamber is incorporated into the experimental platform for XUV energy calibration and time-zero determination. The third component on the platform is a knife-edge apparatus consisting of commercial scalpel blades (Ted Pella) and an XUV photodiode (AXUV100TF400, Opto Diode Co.) for characterization of the XUV beam size at the sample position. Knife-edge measurements yield the horizontal and vertical XUV beam sizes of $97.4\pm0.7$~$\upmu$m and $92.3\pm0.4$~$\upmu$m, respectively. A sample-monitoring camera (Imaging Source) is implemented outside of the chamber to provide real-time visualization of the interaction region, facilitating sample navigation, beam alignment, and optimization of the pump-beam spatial overlap on the calibration gas cell.

\begin{figure}[tbh!]
	\includegraphics[width=\columnwidth]{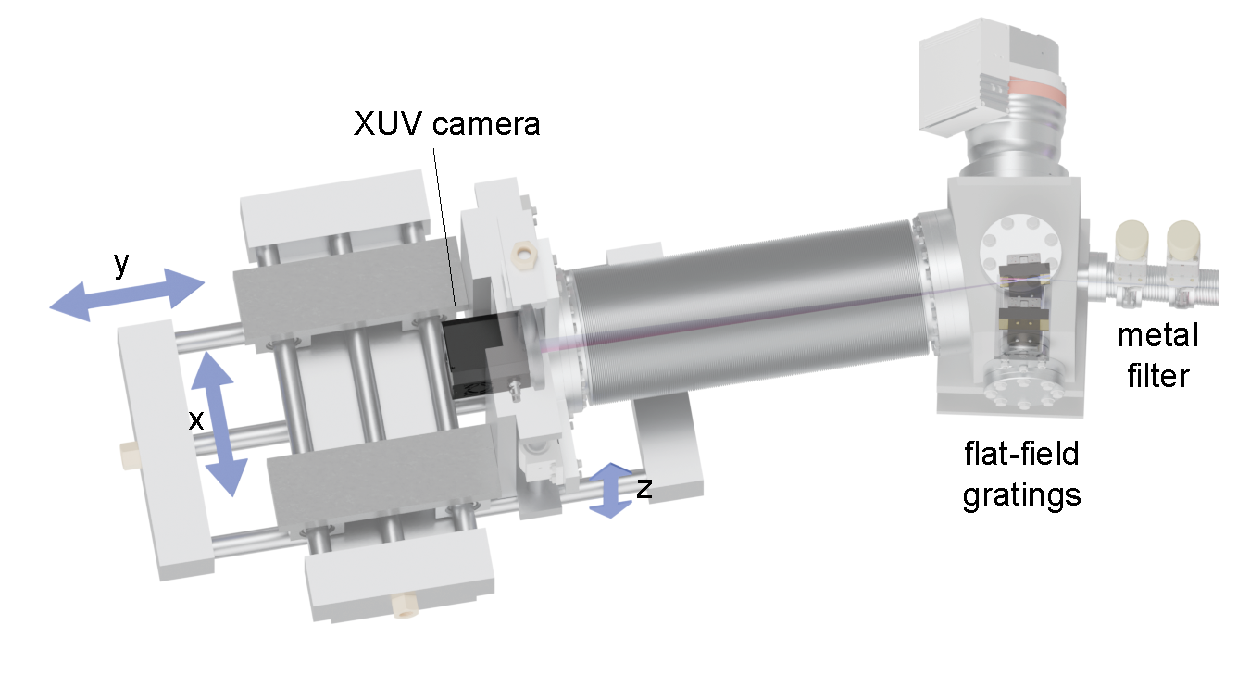}
	\caption{\textbf{Overview of the custom XUV spectrometer.} The transmitted XUV beam passes through interchangeable metal filters mounted on independent VAT gate valves before entering the spectrometer chamber and being dispersed by one of two flat-field gratings mounted on a motorized translation stage. The two gratings provide complementary spectral coverage and spectral resolution, allowing the spectrometer configuration to be tailored to different experimental requirements. The dispersed spectrum is recorded by an XUV CCD detector mounted on a custom three-axis translation stage. Detector motion along the y-axis is used to optimize the spectrometer focus, while translation along the x-axis enables selection of different photon-energy windows. The z-axis is used to optimize the beam position on the detector.} 
\label{fig:6}
\end{figure}

\subsection{XUV spectrometer chamber}
The transmitted XUV beam through the sample then propagates into a homemade XUV spectrometer for spectroscopy measurements. As shown in Fig.~\ref{fig:6}, the spectrometer consists of two filter assemblies mounted on independent manual gate valves (VAT), an interchangeable grating assembly, and an XUV CCD detector (ALEX, Greateyes GmbH). The spectrometer operates under ultrahigh-vacuum conditions with a base pressure on the order of 10$^{-9}$~Torr. The filter assemblies provide additional suppression of the residual NIR driving field and scattered visible light, while also protecting the CCD detector from overexposure. The grating unit includes two flat-field gratings (001-0437 and 001-0660, Hitachi) mounted on a translation stage (Attocube). The former is optimized for broadband spectral acquisition, while the latter provides enhanced spectral resolution for measurements over narrower photon-energy ranges. Unless otherwise specified, the spectrometer characterization and benchmark measurements presented here were acquired using the 001-0660 grating. The dispersed XUV light is recorded by the CCD detector, which is cooled to $-40^{\circ}$C when collecting data. The detector is mounted in a custom-built vacuum enclosure equipped with three-axis translation, enabling optimization of the spectrometer focus ($y$-axis), adjustment of the photon-energy window projected onto the CCD ($x$-axis), and alignment of the dispersed beam on the CCD ($z$-axis). Owing to the large compressive (or tensile) stress exerted by the 8-inch bellow connecting the XUV camera and the chamber that houses the gratings, the three-axis translation stage adopts a particularly sturdy design with 1.25-inch-diameter stainless steel axle to maintain mechanical rigidity. 

\section{Data acquisition and measurement protocols}
Instrument control and data acquisition are implemented through a custom LabVIEW interface that provides centralized control of the beamline and automates common alignment and measurement procedures. To assist beamline alignment, live images from cameras monitoring the HHG and experimental gas cells (Figs.~\ref{fig:3} and \ref{fig:5}) are integrated into the control software. The gas-cell positions are scanned along all three spatial coordinates. Transverse scans maximize XUV transmission through the gas-cell aperture, whereas scans along the beam propagation direction optimize the HHG yield and phase-matching conditions. Following optimization, the centers of the gas cell are registered as reference markers on the live camera images. These markers provide visual references for beamline alignment and sample navigation. 

\subsection{Static XUV measurements}\label{section_reference_for_raster_scans}
For static measurements, XUV transmission spectra are acquired through samples and reference substrates (e.g., silicon nitride TEM windows or copper grids). The corresponding absorption spectra are calculated by 
\begin{align}
    \text{Absorption} = 1 - \frac{I_\text{sample}(E)}{I_\text{blank}(E)},\label{eq:1}
\end{align}
where $I_\text{sample}(E)$ and $I_\text{blank}(E)$ denote the transmitted XUV intensities measured through the sample and reference substrate as a function of photon energy, respectively. This treatment assumes negligible reflection losses, which is generally justified by the near-zero reflectivity of XUV light at normal incidence. To improve spectral quality, both the sample and reference regions are scanned in the transverse and longitudinal directions. Transverse scans average over spatial inhomogeneities in the samples and substrates, while longitudinal scans suppress spectral modulations originating from the intrinsic HHG emission profile through averaging of spectra acquired at different positions along the beam propagation direction \cite{Volkov2019}.

\subsection{Time-resolved XUV measurements}

Time-resolved XUV measurements are conducted by sequential acquisition of pump-off and pump-on spectra at each pump-probe delay. The pump-probe delay is controlled by the 50-cm-long in-vacuum delay stage (Fig.~\ref{fig:3}), enabling measurements spanning sub-fs to 3.3-ns timescales. For a given delay position, the mechanical shutter is alternately closed and opened to acquire a pair of pump-off and pump-on spectra, respectively. The synchronization among the delay stage, mechanical shutter, and XUV CCD camera exposure was controlled through our custom LabVIEW software. %To maximize the detection duty cycle while avoiding pump-induced artifacts, the minimum dead time between successive acquisitions was determined empirically using transient measurements on Ne gas. %As shown in Fig.~\ref{fig:shutter}, dead times shorter than 110~ms produced measurable pump-induced artifacts in the pump-\emph{off} spectra, whereas no discernible artifacts were observed for dead times of 110~ms or longer. Based on this characterization, 
A CCD exposure time of 120~ms was used for transient measurements reported here. %This exposure time provides a balance between photon collection and acquisition efficiency while avoiding unnecessarily long integrations that are susceptible to $1/f$ noise.

%increase susceptibility to slow experimental drift.

%\begin{figure}[tbh!]
	%\includegraphics[scale=0.6]{shutter_exposure_time.pdf}
	%\caption{\textbf{Characterization of the minimum acquisition dead time.} For each tested dead time, the average pre-time-zero spectrum of the nominal pump-off dataset was subtracted from the entire pump-off dataset. Residual transient-like features at the Ne autoionizing resonances are observed for dead times of 90 and 100~ms, whereas no discernible residual signal remains at 110~ms. This measurement establishes the minimum dead time used to optimize the transient acquisition protocol. All traces are vertically displaced for clarity.} 
%\label{fig:shutter}
%\end{figure}

At each delay time point, a pump-on spectrum and the corresponding pump-off spectrum are acquired consecutively. This alternating acquisition sequence is repeated multiple times to improve the signal-to-noise ratio while minimizing the influence of slow experimental drifts. For the measurements presented here, 20 pump-off and 20 pump-on spectra were averaged separately prior to analysis. The transient XUV response is calculated as the normalized transmission change, 
\begin{align}
\Delta I(E,t)=\frac{I_\mathrm{on}(E,t)-I_\mathrm{off}(E)}{I_\mathrm{off}(E)},\label{eq:2}
\end{align}
where $I_\mathrm{on}(E,t)$ denotes the averaged transmitted XUV intensity acquired with the pump beam present at delay time $t$, and $I_\mathrm{off}(E)$ is the transmitted intensity without pump. This interleaved acquisition protocol suppresses low-frequency fluctuations, routinely enabling pre-time-zero noise floors of approximately 0.3--0.7~mOD (see below). To remove long-timescale offsets (e.g., millisecond cumulative heating effects), the averaged pre-time-zero signals are subtracted from the corresponding transient dataset. Time zero is calibrated using NIR-pump-XUV-probe transient measurements on noble gases, typically Ar, He or Ne. The extracted delay time offset is then used as a time zero reference, and the subsequent transient measurements are shifted accordingly. 

\subsection{Temperature-dependent measurements}
\begin{figure}[tbh!]
	\includegraphics[scale=0.6]{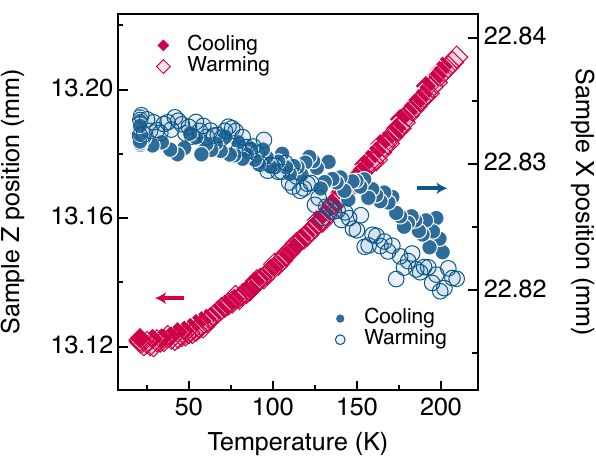}
	\caption{\textbf{Automated sample tracking during temperature cycling.} Sample stage corrections determined by the automated image-registration system during a temperature cycle of NiI$_2$ (see Fig.~\ref{fig:NiI2_static_temp} for XUV spectral trends). The X and Z positions exhibit reproducible cooling and warming trajectories with a small hysteresis, demonstrating reliable compensation of thermally induced sample drift and consistent positioning throughout temperature-dependent measurements.} 
\label{fig:image_registration}
\end{figure}

In addition to static and time-resolved measurements, the c-UBXAS beamline supports automated temperature-dependent XUV spectroscopy. Temperature-dependent measurements are further assisted by a machine-vision-based image registration system that compensates for sample-position drift arising from thermal contraction and expansion during temperature changes. Prior to temperature-dependent measurements, a calibration between image displacement and stage motion is established by translating the sample through known transverse displacements while recording the corresponding pixel shifts in the live camera images. The pixel shifts are determined with sub-pixel precision using image-registration routines implemented in the \textit{scikit-image} Python package\cite{python2014}. During operation, the displacement between the current image and a reference image is calculated and converted into stage corrections through the calibration curve, which are subsequently applied automatically through the motorized translation stages. Following each correction step, the residual displacement relative to the reference image is evaluated and used as feedback for subsequent adjustments until the desired positioning accuracy is achieved. The performance of the image-registration system is illustrated in Fig.~\ref{fig:image_registration}, which shows the stage positions applied during a representative temperature cycle on NiI$_2$. The cooling and warming trajectories closely overlap, demonstrating the reproducibility of the correction procedure and confirming that the same sample region can be reliably revisited throughout repeated temperature sweeps. Residual positioning errors after correction are typically below 1~$\upmu$m, substantially smaller than the $\sim100$~$\upmu$m XUV probe size. This approach enables automated temperature-dependent measurements while maintaining the XUV probe on the same sample region throughout the temperature sweep.

\subsection{Spectral calibration}

XUV photon energy calibration is performed using absorption resonances of He, Ne, and Ar measured with the calibration gas cell integrated into the experimental platform. The resonance energies used for calibration are summarized in Table~\ref{table:calibration}. In addition, the Al $L_{2,3}$ absorption edge introduced by the metal filters provides an internal calibration reference. The identified calibration features are fitted with a three-order polynomial function to establish the pixel-to-energy calibration curve, which is then used to convert detector pixel positions into photon energies for all subsequent measurements.

\begin{table*}[tbh!]
\centering
\caption{Selected atomic levels used for the calibration of the XUV spectrometer \cite{Zong2024}.}
\label{table:calibration}
\begin{footnotesize}
\begin{tabular}{>{\centering\arraybackslash}p{3cm} >{\centering\arraybackslash}p{4cm} >{\centering\arraybackslash}p{4cm} >{\centering\arraybackslash}p{3cm}}
\toprule
\textbf{Noble gas} & \textbf{State} & \textbf{Energy (eV)} & \textbf{Reference} \\
\midrule
\addlinespace[0.5em]
\multirow{3}{*}{\centering He} & $24sp+$ & 64.465(7) & \multirow{3}{*}{Lipsky \& Russek \cite{Lipsky1966}} \\
                         & $23sp+$ & 63.654(6) & \\
                         & $2s2p$   & 60.126(15) & \\
\midrule
\addlinespace[0.5em]
\multirow{5}{*}{\centering Ne} & $2s^12p^66p$ & 47.9650(30) & \multirow{5}{*}{Schulz \textit{et al.} \cite{Schulz1996}} \\
                         & $2s^12p^65p$ & 47.6952(15) & \\
                         & $2s^12p^64p$ & 47.1193(50) & \\
                         & $2s^12p^63p$ & 45.5442(50) & \\
                         & $2p^43s3p$   & 44.9817(50) & \\
\midrule
\addlinespace[0.5em]
\multirow{5}{*}{\centering Ar} & $(^1S)3d(^2D_{5/2})4p$ & 34.994(2)  & \multirow{5}{*}{Madden \textit{et al.} \cite{Madden1969}} \\
                         & $(^1D)3d(^2P_{3/2})4p$ & 34.412(4) & \\
                         & $(^3P)3d(^2D_{3/2})4p$ & 31.624(2)  & \\
                         & $(^3P)4s(^2P_{1/2})5p$ & 31.602(2)  & \\
                         & $(^1D)4s(^2D_{5/2})4p$ & 31.249(2)  & \\
\bottomrule
\end{tabular}
\end{footnotesize}
\end{table*}

\section{Results and discussion}
\subsection{c-UBXAS beamline characterization}

\subsubsection{XUV flux}
\begin{figure}[tbh!]
	\includegraphics[width=\columnwidth]{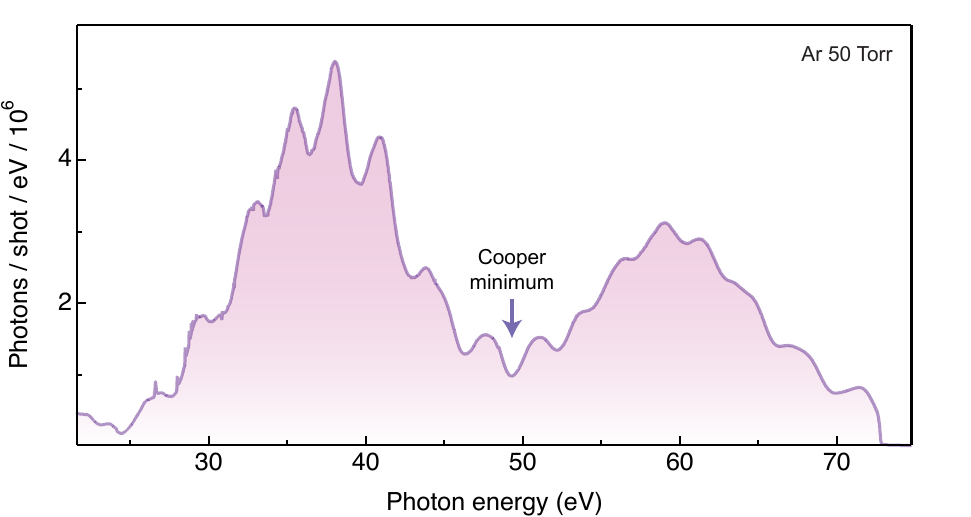}
	\caption{\textbf{HHG spectrum.} Calibrated HHG photon flux density spectrum generated in 50~Torr Ar, providing continuous spectral coverage from 22 to 73~eV. The calibration accounts for the CCD gain and the wavelength-dependent quantum efficiency provided by the detector manufacturer. The spectral features near 26--30~eV originate from reabsorption by the Ar in the generation medium. The broad minimum near 50~eV is attributed to the Ar Cooper minimum \cite{Samson2002}, whereas the sharp cutoff near 73 eV originates from the transmission edge of the 200~nm Al filter.}
\label{fig:hhg}
\end{figure}

Figure~\ref{fig:hhg} shows a representative HHG flux density spectrum generated in a 50~Torr Ar medium through the gas cell. Calibration is performed using the CCD gain and the wavelength-dependent quantum efficiency provided by the detector manufacturer. The resulting spectrum provides continuous spectral coverage from 22 to 73~eV, enabling access to multiple elemental absorption edges within a single spectral frame \cite{Bearden1967}. By integrating the calibrated spectrum recorded on the Greateyes XUV CCD with no sample in the beam path, the photon flux reaching the detector is estimated to be $2\times10^8$~photons per pulse. The resulting flux compares favorably with previously reported HHG-based XUV absorption beamlines \cite{Loh2008,Ryland2018,Ash2023}. Independent XUV photodiode measurements at the sample interaction region (Fig.~\ref{fig:5}) indicate that the XUV pulse energy delivered to the sample is approximately one-millionth of the 1.22~mJ NIR driving pulse energy. This XUV pulse energy value is estimated using the responsivity of the photodiode and corresponds to the standard beamline operating configuration, including transmission through the upstream 200~nm aluminum filter and reflection from the toroidal mirror (Fig.~\ref{fig:4}).

\subsubsection{Spectral resolution}
\begin{figure*}[htb!]
	\includegraphics[scale=0.65]{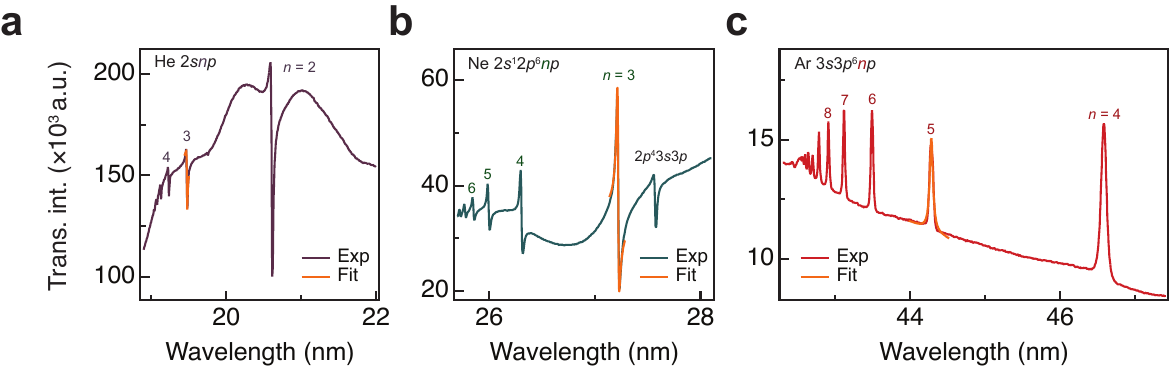}
	\caption{\textbf{Characterization of the XUV spectrometer.} Transmitted XUV spectrum on the He doubly excited states (\textbf{a}), the Ne~$2s^12p^6np$ states (\textbf{b}), and  the Ar~$3s3p^6np$ states (\textbf{c}). Orange curves indicate fits to selected resonances used for resolution analysis. The Ar resonances were fitted using Lorentzian line shapes with a linear background, while the Ne and He resonances were fitted using Fano profiles to account for the asymmetric line shapes arising from continuum coupling. The extracted linewidths were subsequently used to determine the instrumental energy resolution.}
\label{fig:resolution}
\end{figure*}

The spectral performance of the XUV spectrometer is evaluated using well-established spectral features of noble gases. Representative transmission spectra of He, Ne, and Ar are shown in Fig.~\ref{fig:resolution}a--c. These spectral features provide robust references for energy calibration (Table~\ref{table:calibration}) and enable quantitative assessment of the spectrometer spectral resolution. In particular, the resolution is estimated using the He~$2s3p$ doubly-excited states, Ne~$2s2p^63p$ autoionizing state, and Ar~$3s3p^66p$ autoionizing state as references, whose inverse lifetimes are 10~meV, 13~meV, and 12.6~meV, respectively.\cite{Madden1969,stener1995,Domke1996} The extracted FWHMs were 0.013~nm for He, 0.016~nm for Ne, and 0.042~nm for Ar, corresponding to spectral linewidths of 43, 27, and 27~meV, respectively. To isolate the instrumental resolution from the intrinsic linewidth, we apply the standard deconvolution relation $\Delta E_\text{instrum} = \sqrt{(\Delta E_\text{meas})^2-(\Delta E_\text{intrinsic})^2}$, where $\Delta E_\text{instrum}$ represents the instrumental resolution, $\Delta E_\text{meas}$ is the measured linewidth (FWHM), and $\Delta E_\text{intrinsic}$ is the known inverse lifetime width. After correcting for the intrinsic linewidths, the instrumental resolution was determined to be 24--42~meV (0.013--0.037~nm) across the investigated spectral range (22--73~eV).

\begin{figure*}[htb!]
	\includegraphics[scale=0.65]{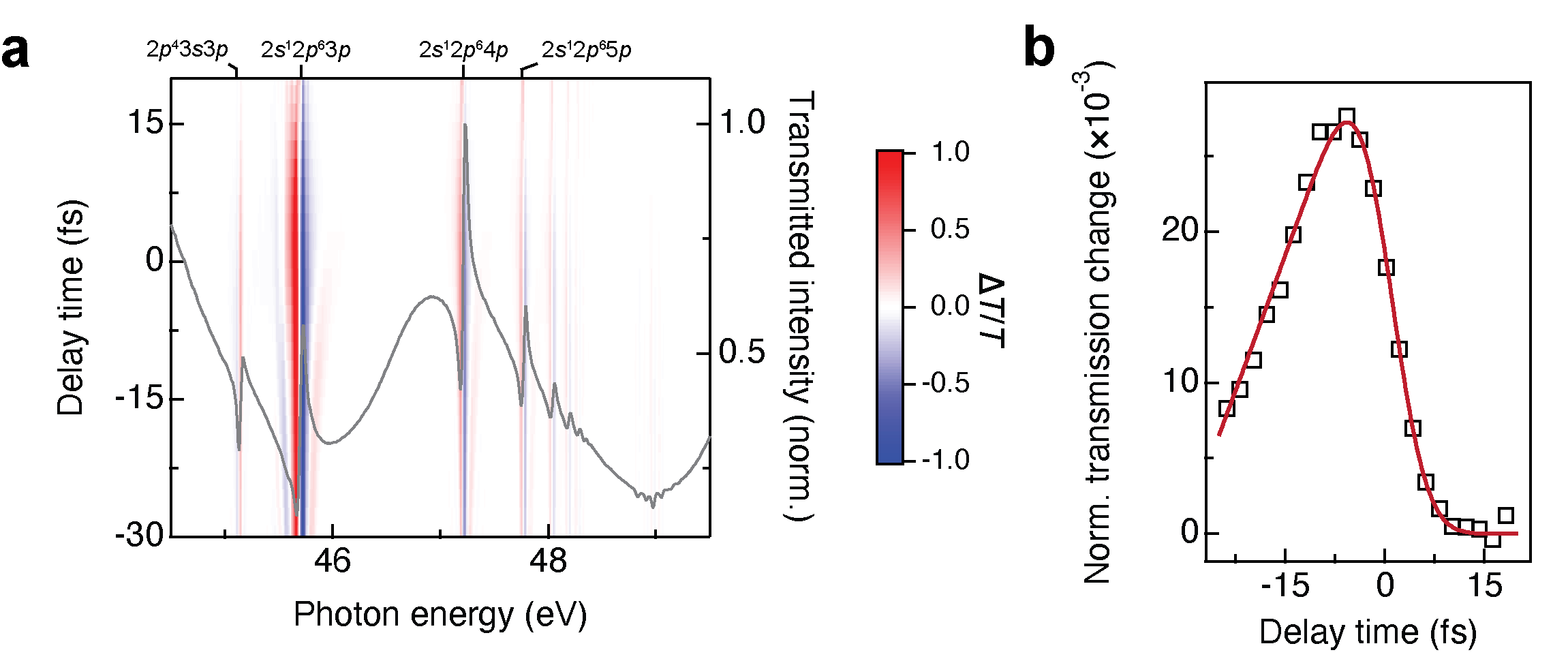}
	\caption{\textbf{Characterization of c-UBXAS temporal resolution.} \textbf{a},~Transient transmission spectrogram of Ne autoionizing states. The equilibrium transmitted XUV spectrum was plotted against the right axis as a reference for the corresponding states, which are labeled at the top. Positive (negative) time delay means that XUV pulse arrives after (before) the NIR pump pulse. This measurement used 30~Torr of Ne, measured before it exited the gas cell. \textbf{b},~Transient signal of the NIR dressing Ne $2s^12p^63p$ side band region (45.8--46.3~eV) extracted from \textbf{a}. Red curve is a fit to Eq.~\eqref{eq:fit_erfexp}.}
\label{fig:irf}
\end{figure*}
\subsubsection{Temporal resolution}
The temporal response of the setup is estimated by performing transient XUV measurement on Ne. Figure~\ref{fig:irf}a shows the transient response of the series of Ne $2s^12p^6np$ autoionizing states, in which the temporal trace integrated over the NIR dressing sideband region (45.8--46.3~eV) serves as a proxy for the effective temporal response of the beamline \cite{Jager2017}. The resulting trace was fitted using the phenomenological model function \cite{Hellmann2012,Moore2016,Zong2021}
\begin{align}
    \Delta{I}(t) = \frac{1}{2}\left[1 + \mathrm{Erf}\left(\frac{2\sqrt{\ln 2}(t-t_0)}{\omega}\right)\right]~~~~~~~~~~~\notag\\
    \cdot \left(I_\infty + I_0 e^{-(t-t_0)/\tau}\right),\label{eq:fit_erfexp}
\end{align}
where $\omega$ represents the width of the error-function rise, $I_0$ stands for the maximum intensity change, $I_\infty$ is the value of $\Delta I$ at long time delays, $\tau$ denotes the characteristic relaxation time to the quasi-equilibrium, and $t_0$ is associated with the relative arrival time of pump and probe when $\Delta I = (I_0+I_\infty)/2$. As shown in Fig.~\ref{fig:irf}b, the fit yields the width of the error function $\omega = 9.8\pm0.7$~fs, providing an upper-bound estimate of the instrument response function \cite{Zurch2017}.

\subsubsection{Long-term stability}
\begin{figure}[htb!]
	\includegraphics[width=\columnwidth]{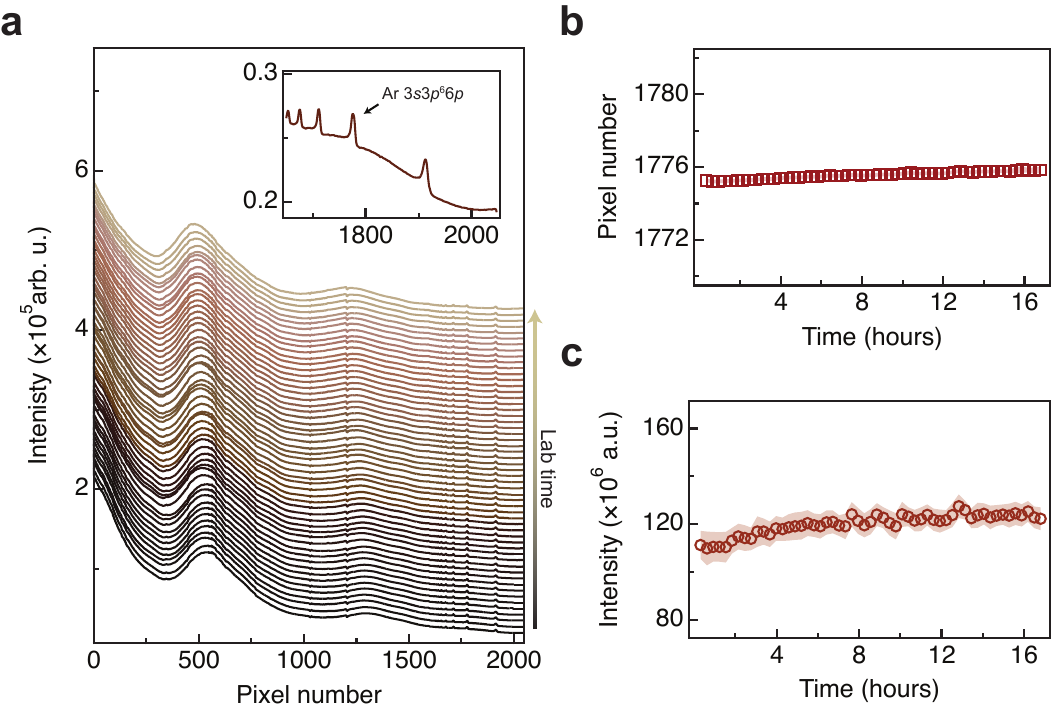}
	\caption{\textbf{Long-term stability of the c-UBXAS beamline.} \textbf{a},~Raw XUV spectra acquired during a 16-hour continuous measurement on an empty silicon nitride membrane. Consecutive spectra are vertically offset for clarity and ordered according to laboratory acquisition time. The inset highlights the Ar~$3s3p^66p$ autoionizing reabsorption feature used as an internal spectral marker. \textbf{b},~Extracted peak position of the Ar~$3s3p^66p$ autoionizing resonance as a function of lab time, showing a positional uncertainty of 0.2~pixel over the measurement period. The shaded region represents 1~s.d. peak fitting uncertainties. \textbf{c},~Integrated XUV intensity as a function of lab time. The shaded region denotes the 1~s.d. variation about the mean intensity.}
\label{fig:stability}
\end{figure}

We next evaluate the long-term beam pointing and intensity stability of the XUV spectrometer. 
For this purpose, we monitored the transmitted XUV intensity through an empty silicon nitride TEM window, a substrate commonly used for XUV spectroscopy measurements (Fig.~\ref{fig:stability}). 
%For this purpose, a dummy pump-probe scan was performed on an empty silicon nitride TEM window, a substrate commonly used for XUV spectroscopy measurements (Fig.~\ref{fig:stability}). The scan covered pump-probe delays from $-200$~fs to 2~ps with equal spacing of 20~fs and a 120~ms CCD exposure time. 
The Ar $3s3p^66p$ autoionizing reabsorption feature from the HHG source was used as an internal spectral marker. Over 16~hours of continuous operation, the peak position is found to be stable within an uncertainty of 0.2~pixel, corresponding to a spectral shift of approximately 1~meV. %We note that the energy shift associated with a given pixel displacement depends on the selected spectral window. 
In addition, the integrated signal across the entire spectra exhibited an rms fluctuation of 3.6\%. These results demonstrate the long-term source stability required for extended XUV spectroscopy measurements.

\section{Static and time-resolved benchmark measurements on multiferroic NiI$_\text{2}$}
To benchmark the capabilities of the cryogenic beamline, we performed temperature-dependent XUV spectroscopy measurements on NiI$_2$. NiI$_2$ is a layered transition metal halide consisting of edge-sharing [NiI$_6$] octahedra separated by van der Waals gaps. The material undergoes a series of phase transitions upon cooling, first transitioning from a paramagnetic to an antiferromagnetic (AFM) state below 75~K and subsequently undergoing a rhombohedral-to-monoclinic structural distortion below 59~K. The latter transition is accompanied by the evolution of the AFM order into a proper-screw helical magnetic structure. Given the coupling between its magnetic and lattice degrees of freedom, it is classified as an improper type-II multiferroic. Recent studies have further shown that this multiferroic state persists down to the monolayer limit, highlighting NiI$_2$ as a promising candidate for low-dimensional spintronic and multifunctional electronic devices \cite{Ju2021, Song2022, Lebedev2023, Song2025, Gish2024, Lebedev2024}. To realize these opportunities, it is essential to understand the electronic structure responsible for multiferroicity and how it evolves across phase transitions and under nonequilibrium conditions \cite{Fiebig2016}. In particular, studies have suggested that the multiferroicity is closely linked to strong hybridization between Ni 3$d$ and I 5$p$ states \cite{Kurumaji2013, Riedl2022, Kapeghian2024}. To understand this element-specific orbital characters and their dynamics, ultrafast broadband XUV spectroscopy is uniquely suited to directly visualize such interactions.

\begin{figure}[t!]
	\includegraphics[width=0.8\columnwidth]{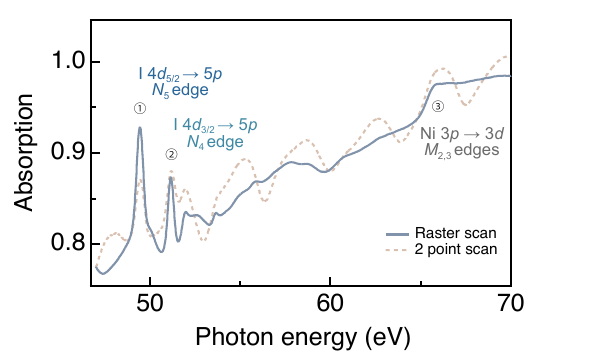}
	\caption{\textbf{Static XUV absorption spectrum of NiI$_\text{2}$.} Spectra obtained using a two-point measurement and a raster scan are shown for comparison. Features \circled{1} and \circled{2} correspond to the I~$N_5$ ($4d_{5/2}\rightarrow5p$) and I~$N_4$ ($4d_{3/2}\rightarrow5p$) absorption edges, respectively, while feature \circled{3} originates from the Ni~$M_{2,3}$ ($3p\rightarrow3d$) absorption edges. The raster scan reduces spurious spectral structure and yields a more representative absorption spectrum of the sample. See Sec.~\ref{section_reference_for_raster_scans} for details of the measurement protocol.}
\label{fig:NiI2}
\end{figure}

Figure~\ref{fig:NiI2} shows a static XUV absorption spectrum of NiI$_2$ measured at 300~K using the raster-based acquisition protocol described in Sec.~\ref{section_reference_for_raster_scans}. Compared with a simplified two-point measurement, the raster-based approach suppresses HHG source-induced spectral modulations while preserving the underlying spectral features. The resulting spectrum spans both iodine and nickel absorption regions in a single spectral frame, providing simultaneous access to electronic states relevant to the Ni 3$d$ and I 5$p$ hybridization in the material. In the iodine region, two main absorption peaks, labeled as \circled{1} and \circled{2}, are observed and attributed to I $N_{4,5}$ transitions from the I $4d$ spin-orbit split core levels to unoccupied I $5p$ states near the Fermi level \cite{Kobayashi2019}. The assignment is supported by the $\sim1.7$~eV separation between the two features, matching the spin-orbit splitting of the I $4d_{5/2}$ and 4$d_{3/2}$ core levels \cite{Brown1970}. On the other hand, a broad absorption feature, labeled \circled{3}, spanning 65--67~eV is assigned to the Ni $M_{2,3}$ edge, arising from $3p\rightarrow3d$ transitions, consistent with previous synchrotron measurements \cite{Wang2013}.

\begin{figure*}[htb!]
	\includegraphics[scale=0.5]{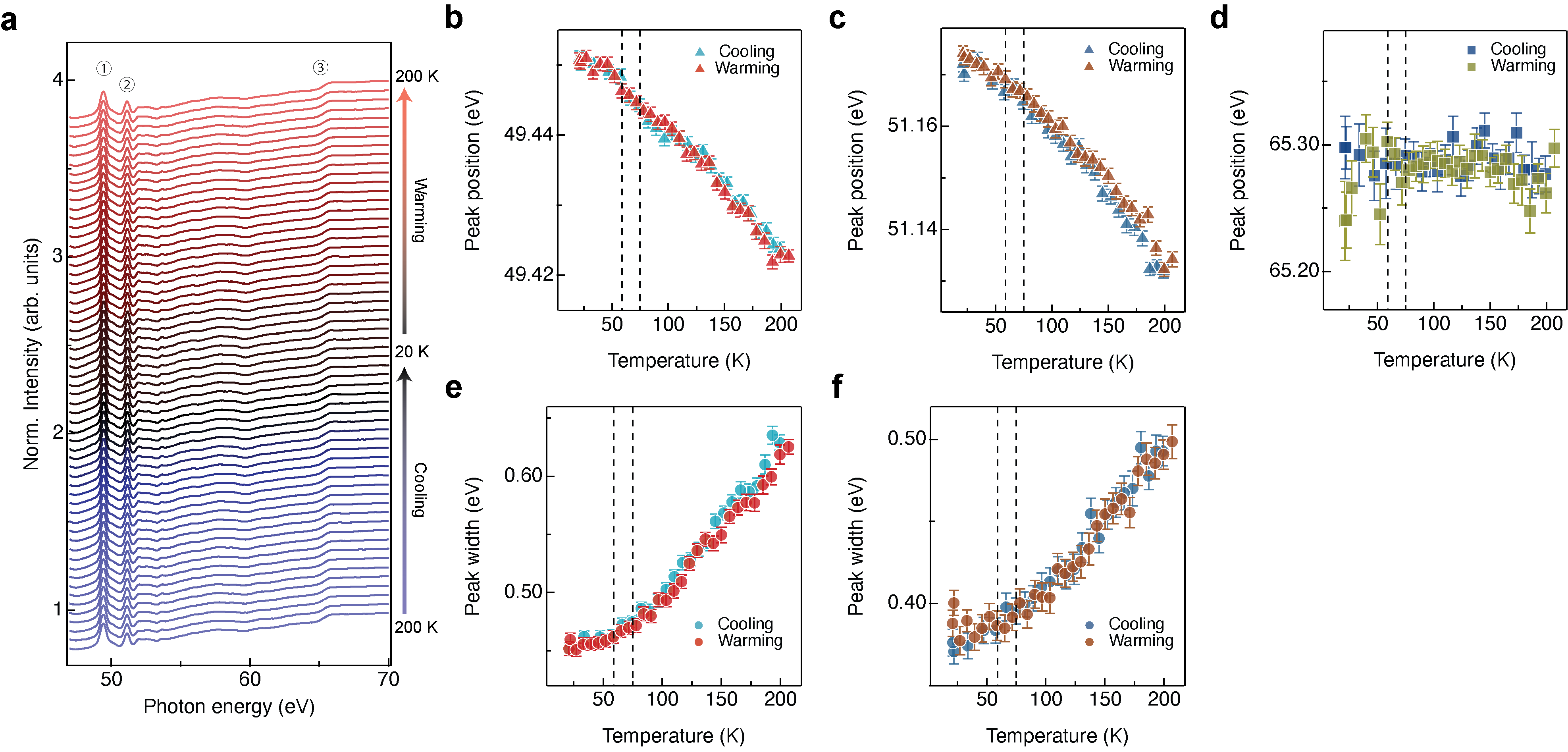}
	\caption{\textbf{Temperature-dependent XUV absorption spectra of NiI$_\text{2}$.} \textbf{a},~Normalized absorption spectra acquired during cooling (blue) and subsequent warming (red) during a 200--20--200~K thermal cycle. Features~\circled{1} and \circled{2} correspond to the I $N_5$ and I $N_4$ absorption edges, respectively, while feature \circled{3} originates from the Ni $M_{2,3}$ edges. Automated image-registration correction (Fig.~\ref{fig:image_registration}) was employed throughout the measurement to compensate thermally induced sample drift and maintain the XUV probe on the same sample region. \textbf{b--d},~Temperature dependence of the peak positions of features \circled{1}--\circled{3} during cooling (filled symbols) and warming (open symbols). The peak positions of the iodine features were determined from Lorentzian fits with a constant background, whereas the position of the nickel feature was estimated from the inflection point associated with the shoulder profile. \textbf{e},\textbf{f},~Temperature dependence of the linewidths extracted from the Lorentzian fits to features \circled{1} and \circled{2}. Vertical dashed lines indicate the reported phase transition regime of NiI$_2$. Error bars represent 1~s.d. fitting uncertainties.}
\label{fig:NiI2_static_temp}
\end{figure*}

Having established the spectral assignments, we next investigate their temperature dependence across the phase transitions of NiI$_2$. Leveraging the combined cryogenic measurement platform and image-registration procedure, we tracked the evolution of the NiI$_2$ XUV absorption spectrum during a continuous 200--20--200~K thermal cycle, with spectra acquired at temperature intervals of approximately 6--8~K. As summarized in Fig.~\ref{fig:NiI2_static_temp}, the resulting dataset spans both phase transitions and demonstrates the stability and reproducibility of the setup over extended temperature excursions. To quantify the temperature-dependent spectral evolution, the positions and widths of the absorption features were extracted. Both iodine features exhibit gradual energy shifts upon cooling (Fig.~\ref{fig:NiI2_static_temp}b,c), while the nickel feature shown in Fig.~\ref{fig:NiI2_static_temp}d remains largely unchanged. The linewidths of the iodine peaks also exhibit a systematic temperature dependence, including a subtle change in slope near the reported transition-temperature regime of NiI$_2$ (Fig.~\ref{fig:NiI2_static_temp}e,f). While a detailed interpretation is beyond the scope of the present work, the observation highlights the sensitivity of the cryogenic XUV platform to subtle spectral changes involved in low-energy phase transitions in quantum materials. 

\begin{figure*}[htb!]
	\includegraphics[scale=0.6]{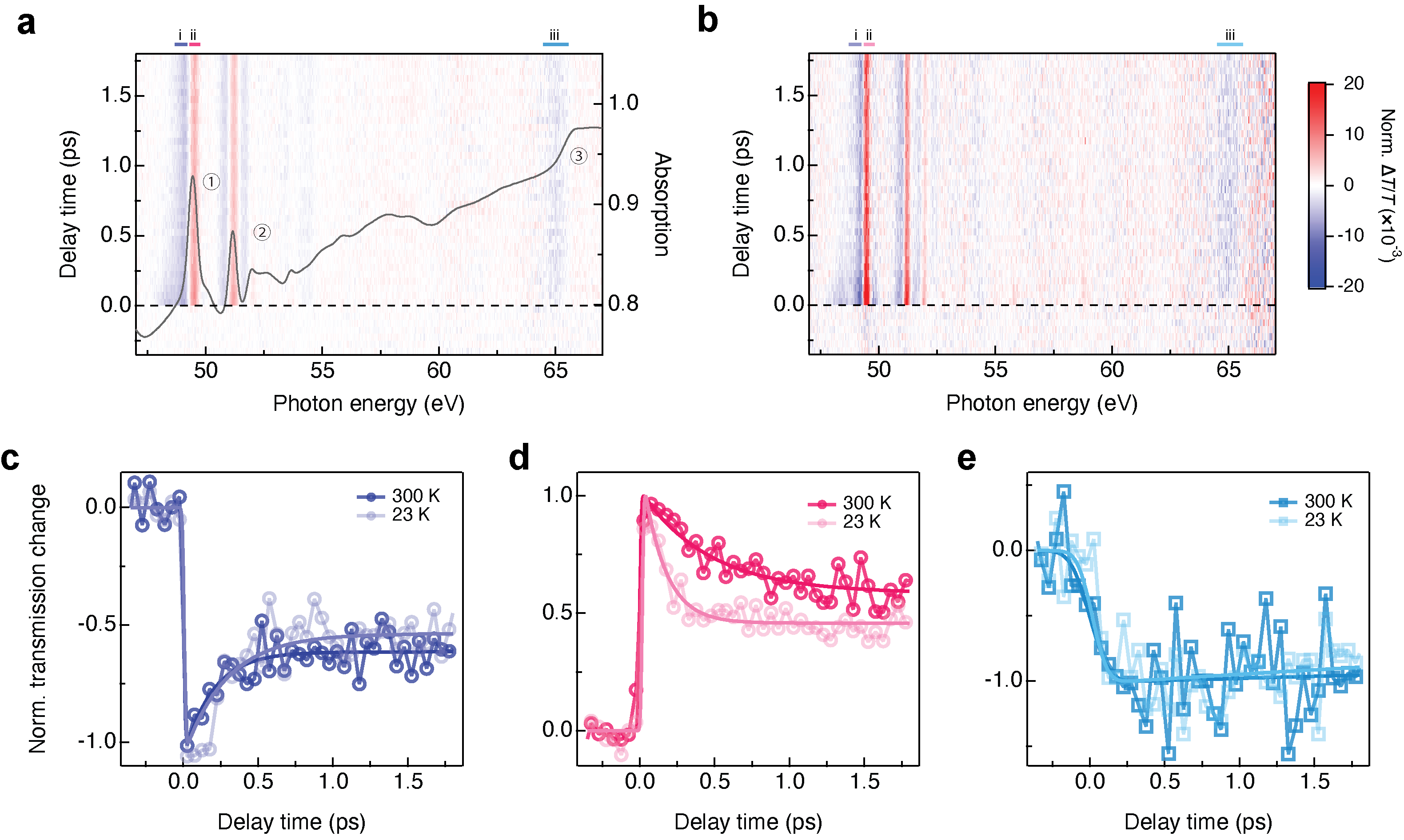}
	\caption{\textbf{Temperature-dependent transient XUV absorption measurements of NiI$_\text{2}$.} \textbf{a},\textbf{b},~Spectrally resolved transient transmission changes measured at (\textbf{a}) 300~K and (\textbf{b}) 23~K following photoexcitation at a fluence of 2.0~mJ/cm$^2$. The static absorption spectrum shown in gray in \textbf{a} is reproduced from Fig.~\ref{fig:NiI2} and plotted against the right-hand axis for reference. Colored bars labeled i--iii on top denote the three spectral integration windows used for the transient analyses shown in \textbf{c}--\textbf{e} corresponding to 48.67--49.23~eV (i), 49.30--49.78~eV (ii), and 64.46--65.55 eV~(iii), respectively. \textbf{c}--\textbf{e},~Transient traces obtained by integrating the transient signal within the corresponding spectral windows. Dark and light symbols denote measurements at 300~K and 23~K, respectively, while solid curves show fits to the transient response using the Eq.~\eqref{eq:fit_erfexp}. The transient signals are normalized to their respective extrema to clearly identify the relaxation timescales.}
\label{fig:NiI2_dynamics_temp}
\end{figure*}

We next demonstrate the ultrafast capabilities of the cryogenic XUV beamline through pump-probe measurements on NiI$_2$. An NIR pump pulse with incident fluence of 2~mJ/cm$^2$ excites carriers across the 1.2~eV bandgap of the material \cite{chen1993}, and the resulting electronic dynamics are monitored through the iodine and nickel regions. Figure~\ref{fig:NiI2_dynamics_temp}a,b shows transient transmission change acquired at 300~K and 23~K, corresponding to temperatures above and below the magnetic and structural phase transitions. At both temperatures, distinct positive and negative transmission changes are observed within the iodine region, whereas only a negative change is resolved at the Ni $M_{2,3}$ edge within the current signal-to-noise level. The experimental noise floor, estimated from the pre-time-zero regions of both datasets, ranges from $7.8\times10^{-4}$ to $1.6\times10^{-3}$ in normalized transmission change (corresponding to 0.3--0.7~mOD). These values are comparable to those reported for state-of-the-art XUV transient spectroscopy measurements \cite{Geneaux2021, Gutberlet2023}. 

The temporal evolution of the photoinduced spectral features was further analyzed by integrating over selected spectral windows~i--iii corresponding to the negative and positive features of the I~$N_{5}$ edge and the Ni~$M_{2,3}$ edges. At 300~K, the extracted traces at different absorption edges exhibit distinct temporal evolutions. Specifically, the positive and negative iodine features have a very fast initial response, followed by relaxations on the timescales of $\tau_{\text{red,300~K}}=460\pm40$~fs and $\tau_{\text{blue,300~K}}=210\pm30$~fs [see Fig.~\ref{fig:NiI2_dynamics_temp}c,d and Eq.~\eqref{eq:fit_erfexp}]. By contrast, as shown in Fig.~\ref{fig:NiI2_dynamics_temp}e, the Ni $M_{2,3}$ response develops more gradually with a characteristic rise time $\omega_{\text{Ni,300~K}}=240\pm120$~fs. As a result, the iodine signals begin to recover while the nickel feature is still rising, with the two responses crossing at early times. This crossover may indicate a sequential process involving iodine- and nickel-derived electronic states. Upon cooling to 23~K, the characteristic timescales change to $\tau_{\text{red,23~K}}=150\pm10$~fs, $\tau_{\text{blue,23~K}}=190\pm20$~fs, and $\omega_{\text{Ni,23~K}}=190\pm90$~fs. The most pronounced change is observed for the positive iodine transient feature (red curves in Fig.~\ref{fig:NiI2_dynamics_temp}d), whose relaxation time decreases by approximately a factor of three compared to 300~K. The selective temperature dependence of the iodine response is suggestive of changes in ligand-mediated electronic relaxation pathways associated with the strongly covalent Ni-I electronic structure reported for NiI$_2$ \cite{Son2022,Occhialini2024}, motivating future investigations of the underlying microscopic mechanism. Importantly, these measurements underscore the need for access to cryogenic environments to disentangle element-specific dynamics across phase transitions, a long-standing experimental challenge that c-UBXAS is uniquely positioned to address.

\begin{table}[h!]
\caption{\label{tab:summary} Summary of the measured performance metrics of the c-UBXAS beamline. Values are obtained from the characterization measurements described in the text and benchmark static and time-resolved experiments on NiI$_2$.}
\begin{ruledtabular}
\begin{tabular}{lc}
Metric &Value\\
\hline 
\addlinespace[0.5em]
XUV photon flux & 2$\times10^8$~photons/pulse\\
\addlinespace[0.5em]
XUV probe size & $97.4(7)\times92.3(4)~\upmu\mathrm{m}^2$\\
\addlinespace[0.5em]
Photon energy range & 22--73~eV \\
\addlinespace[0.5em]
Energy resolution & 24--42~meV (0.013--0.037~nm)\footnote{Determined from Ar and He reference features at 29~eV ($44$~nm) and 64~eV ($19$~nm), yielding instrumental resolutions of 24~meV (0.037~nm) and 42~meV (0.013~nm), respectively.}\\
\addlinespace[0.5em]
Temporal resolution & $9.8\pm0.7$~fs\\
\addlinespace[0.5em]
Pump-probe delay range & sub-fs to 3.3~ns\\
\addlinespace[0.5em]
Temperature range & 20--350~K\\
\addlinespace[0.5em]
Noise floor & 0.3--0.7~mOD\\
\addlinespace[0.5em]
Beam pointing stability & 0.2~pixel\footnote{Measured during 16 hours of continuous operation using the Ar~$3s3p^66p$ autoionizing reabsorption feature as an internal spectral marker.}\\
\addlinespace[0.2em]
Flux stability & 3.6\%~rms$^\text{b}$\\
\end{tabular}
\end{ruledtabular}
\end{table}

\section{Conclusion and outlook}
We present a cryogenic ultrafast broadband XUV absorption (c-UBXAS) beamline tailored for investigations of quantum materials under both equilibrium and nonequilibrium conditions. The beamline combines broadband XUV spectroscopy, automated image registration, and cryogenic sample control, enabling temperature-dependent measurements down to 20~K and pump--probe experiments spanning sub-femtosecond to nanosecond timescales. We summarize the key performance metrics discussed in this article in Table~\ref{tab:summary}. Benchmark measurements on NiI$_2$ further reveal spectral shifts and linewidth changes across the phase transition regime, together with the distinct ultrafast dynamics at the iodine and nickel absorption edges. These results demonstrate the capability of the beamline to track subtle spectral evolution across phase transitions and to perform element-resolved measurements of ultrafast electronic dynamics. 

Looking ahead, an exciting direction is the incorporation of X-ray magnetic circular dichroism (XMCD) capabilities, extending the c-UBXAS platform to spin-resolved measurements \cite{Siegrist2019}. Such measurements would allow direct access to element-resolved spin dynamics, angular momentum transfer, and short range magnetic correlations. These capabilities could provide critical insights into quantum materials where precursor magnetic states and fluctuating spin correlations are believed to contribute to the emergence of long-range magnetic order \cite{Wu2024,Lopez2022,Alfonsov2021}, while also enabling studies of nonequilibrium angular momentum transfer processes, including ultrafast manifestations of the Einstein–de Haas effect \cite{Hennecke2019, Dornes2019} and light-induced magnetic states in van der Waals magnets \cite{Ilyas2024,Sharma2026}.

Beyond XMCD, polarization control of the XUV probe would also enable X-ray linear dichroism measurements, providing sensitivity to anisotropic electronic, orbital, and ferroic order parameters. With reduced XUV probe sizes through improved focusing or nanofocusing optics \cite{Koch2018}, such capabilities could facilitate spatially resolved studies of ferroic domains and domain-wall dynamics \cite{Chang2026}. Looking further ahead, nanofocused XUV spectroscopy integrated with nonlinear optical excitation schemes, such as XUV second-harmonic generation, may offer enhanced sensitivity to buried interfaces and interfacial symmetry breaking \cite{Helk2021}. This sensitivity opens new avenues for investigating emergent superstructures in quantum materials, including perovskite oxide superlattices \cite{Yadav2016, Hong2017, Das2019} and moiré heterostructures \cite{Qi2025, Zhang2022}, where interfacial interactions play a central role in determining their collective properties.

We envision the c-UBXAS to serve as a powerful platform for uncovering the microscopic interactions underlying emergent phases in highly nonequilibrium quantum materials \cite{Zong2023}.\\\\\\\\\\

\section{Acknowledgments}
\color{black}
We thank Dmitry Lebedev, Thomas Wei Song, and Professor Mark C. Hersam at Northwestern University for providing NiI$_{2}$ samples used in this work.
We thank Holger~Oertel, William~Windsor, Laurenz~Rettig, Ralph~Ernstorfer, and Martin~Wolf from Fritz Haber Institute  (FHI) Berlin and technical staff at FHI for support and discussion that enabled the construction of the cryogenic endstation in the c-UBXAS instrument.
We thank Doug Scudder of the College of Chemistry Machine Shop at the University of California, Berkeley, for fabricating the custom vacuum enclosure for the XUV CCD camera.
We thank Dr.~Marcus~Goetz at MRC Systems GmbH for providing the STEP file of the piezoelectric steering mirror mount used in the beamline rendering.
% Alfred's acknowledgement
A.Z. acknowledges support from the Miller Institute for Basic Research in Science (instrumentation), the National Science Foundation (NSF-DMR~2247363) (data acquisition), and the U.S. Department of Energy, Office of Basic Energy Sciences under award No.~DE-SC0026202 (data analysis and manuscript writing).
% Sheng-Chih's acknowlegement
S.-C.L. acknowledges support by the Berkeley-Taiwan Fellowship and the National Science Foundation (NSF-DMR 2247363).
% Emma's and Bailey's acknowledgement
E.B. and B.R.N. acknowledge support from the National Science Foundation Graduate Research Fellowships Program under Grant Nos.~DGE~2146752 and 1752814, respectively.
% Marcus's acknowledgement
M.H. acknowledges funding by the National Science Foundation (NSF-REU EEC-1852537). 
% Shuaiwei's acknowledgement
S.P. acknowledges support from the Rose Hills Foundation and the Air Force Office of Scientific Research (AFOSR, FA9550-25-1-0339).
% Jackson's acknowledgement
J.M. acknowledges support from the Department of Energy (DE-SC0024123).
% NSF's disclaimer:
Any opinions, findings, and conclusions or recommendations expressed in this material are those of the authors and do not necessarily reflect the views of the National Science Foundation. 
% Michael's acknowledgement
M.W.Z. acknowledges funding by the W.~M. Keck Foundation, Laboratory Directed Research and Development Program at Berkeley Lab (107573 and 108232), the Rose Hills Foundation, the Camille and Henry Dreyfus Foundation, the Max Planck Society, and the Hellman Fellows Fund which enabled the building of the c-UBXAS instrument. M.W.Z. further acknowledges support by the Department of Energy (DE-SC0024123). \\
\color{black}

\section{Author declarations}
\subsection{Conflict of interest}
The authors have no conflicts to disclose.
\subsection{Author contributions}
Sheng-Chih Lin and Alfred Zong contributed equally to this work.\\

\noindent{\textbf{Sheng-Chih Lin}}: Methodology (equal); Investigation (lead); Writing - original draft (lead); Data curation (lead); Visualization (lead); Formal analysis (lead); Validation (lead). Writing – review $\&$ editing (lead).
\textbf{Alfred Zong}: Methodology (lead); Investigation (lead); Writing - original draft (supporting); Data curation (lead); Visualization (supporting); Formal analysis (lead); Validation (equal). Writing – review $\&$ editing (equal).
\textbf{Emma Berger}: Methodology (equal); Writing – review $\&$ editing (supporting).
\textbf{Bailey R. Nebgen}: Methodology (supporting); Writing – review $\&$ editing (supporting).
%\textbf{Dmitry Lebedev}: Investigation (supporting);  Writing – review $\&$ editing (supporting).
%\textbf{Thomas Wei Song}: Investigation (supporting).
\textbf{Marcus Hui}: Methodology (supporting); Visualization (supporting); Writing – review $\&$ editing (supporting).
\textbf{Shuaiwei Pan}: Methodology (supporting); Visualization (supporting); Writing – review $\&$ editing (supporting).
\textbf{Jackson McClellan}: Methodology (supporting); Writing – review $\&$ editing (supporting).
%\textbf{Mark C. Hersam}: Supervision (supporting).
\textbf{Michael W. Zuerch}: Supervision (lead); Funding acquisition (lead); Conceptualization (lead); Writing – review $\&$ editing (equal).

\section{Data availability}
The data that support the findings of this study will be uploaded to a public repository upon peer-review acceptance. Until then, the data are available from the corresponding author on reasonable request.
%The data that support the findings of this study are available from the corresponding author upon reasonable request.

%\clearpage
%\newpage
\section{References}
%\bibliography{main}
%aipnum4-2.bst 2019-01-14 (MD) hand-edited version of apsrev4-1.bst

\end{document}